\documentclass[aip,graphicx,reprint]{revtex4-1}

\usepackage{graphicx}
\usepackage{dcolumn}
\usepackage{amssymb} 

\begin{document}





\title{Room temperature ballistic transport in InSb quantum well nanodevices} 








\author{A. M. Gilbertson}
\email[Email:]{adam-maurick.gilbertson05@ic.ac.uk}
\affiliation{Blackett Laboratory, Imperial College, London SW7 2BZ, UK}

\author{A. Korm\'{a}nyos}
\affiliation{Department of Physics, Lancaster University, Lancaster LA1 4YB, UK}

\author{P. D. Buckle}
\affiliation{Department of Physics and Astronomy, Cardiff University, Cardiff CF24 CAA, UK}

\author{M. Fearn}
\affiliation{Dept. of Physics, Lancaster University, Lancaster LA1 4YB, UK}

\author{T. Ashley}
\affiliation{ School of Engineering, University of Warwick, Coventry CV4 7AL, UK}

\author{C. J. Lambert}
\affiliation{Department of Physics, Lancaster University, Lancaster LA1 4YB, UK}

\author{S. A. Solin}
\affiliation{Center for Material Innovations and Department of Physics, Washington University in St. Louis, St.Louis, MO-63130, USA}
\affiliation{Blackett Laboratory, Imperial College, London SW7 2BZ, UK}

\author{L. F. Cohen}
\affiliation{Blackett Laboratory, Imperial College, London SW7 2BZ, UK}





\date{\today}

\begin{abstract}
We report the room temperature observation of significant ballistic electron transport in shallow etched four-terminal mesoscopic devices fabricated on an InSb/AlInSb quantum well (QW) heterostructure with a crucial partitioned growth-buffer scheme. Ballistic electron transport is evidenced by a negative bend resistance signature which is quite clearly observed at 295 K and at current densities in excess of 10$^{6}$ A/cm$^{2}$. This demonstrates unequivocally that by using effective growth and processing strategies room temperature ballistic effects can be exploited in InSb/AlInSb QWs at practical device dimensions.

\end{abstract}


\maketitle 

Harnessing ballistic transport effects in low-dimensional structures at room temperature (RT) is a promising avenue for developing novel functionality in nanoelectronic devices for applications including, logic circuits, biosensing and high-density data storage. Carbon-based systems such as carbon nanotubes (CNTs)\cite{Javey2003,Frank1998} and graphene\cite{Mayorov2011} have received considerable attention owing to their extraordinarily long mean free path ($l_{0}$) at RT ($<$ 50 $\mu$m in CNTs) and high current carrying capability, but the realization of very-large-scale-integration compatibility remains a fundamental challenge. In this respect, high mobility III-V semiconductors are technologically relevant. Several groups\cite{Song2001,Song2001b,Hieke2000} have explored novel ballistic switching and rectifying concepts in InGaAs/InP quantum wells (QWs) where $l_{0} \approx$ 150 nm at 295 K. The operating efficiency of such devices is closely linked to the ratio of $l_{0}$ to the critical device dimension and is limited to $\approx 20\%$ due to the small value of $l_{0}$ in such systems. Electron mobilities of $\mu_{e} \approx 45,000$ cm$^{2}$/Vs are routinely achieved in InSb/AlInSb QWs at 295 K,\cite{Orr2008} the largest reported of all III-V systems. For a typical electron density $n_{e} \approx 6$ x 10$^{11}$ cm$^{-2}$ this corresponds to $l_{0} = \hbar k_{F}\mu_{e}/e \approx $ 550 nm (where $k_{F} = (2\pi n_{e})^{1/2}$ is the Fermi wavevector). Considerable advantages would be afforded by pursuing such device concepts in this system, however to-date the RT operation of InSb QW nanodevices has been inhibited by excessive growth-buffer layer leakage currents.\cite{Goel2004}

In this letter we report the magnetotransport properties of mesoscopic devices fabricated on an InSb/AlInSb QW with a partitioned buffer layer (PBL) scheme\cite{Gilbertson2011b} designed to suppress the parasitic leakage, that demonstrate remarkably clear ballistic transport at 295 K as a result.

The sample used is a 15-nm modulation doped InSb/Al$_{x}$In$_{1-x}$Sb QW  grown by MBE onto a GaAs (001) substrate with a PBL scheme (growth details are found elsewhere).\cite{Gilbertson2011b} A 15-nm pseudomorphic Al$_{0.3}$In$_{0.7}$Sb layer was incorporated 300 nm below the QW to provide a potential barrier to electrons and holes, thermally generated in the bulk of the buffer layer, from diffusing to the Ohmic contact region. In this way the effective electrical thickness of the buffer layer is reduced from 3 $\mu$m to 300 nm. The electron density and mobility of the QW at 295 K are $n_{e}$ = 7.31x10$^{11}$ cm$^{-2}$ and $\mu_{e}$ = 41,500 cm$^{2}$/Vs ($l_{0}$ = 586 nm), as deduced from high magnetic field measurements on 40-$\mu$m-wide (reference) Hall bridges.\cite{Gilbertson2011b} Four-terminal mesoscopic cross structures of various sizes were fabricated using $e$-beam and optical lithography and shallow (100 nm etch depth) reactive ion etching (RIE) in a CH$_{4}$/H$_{2}$ (1:8) gas mixture. Magnetotransport measurements were performed in perpendicular magnetic fields (B) up to 7.5 T at various temperatures using standard AC and DC measurement techniques. The sidewall depletion width, $w_{dep}$, was estimated from the dependence of the two-terminal conductance (G$_{2T}$) (B = 0) of several devices on the physical lead width at 160 K to be $w_{dep} \approx$ 120 nm (not shown).

\begin{figure}
\includegraphics[width=7.6cm]{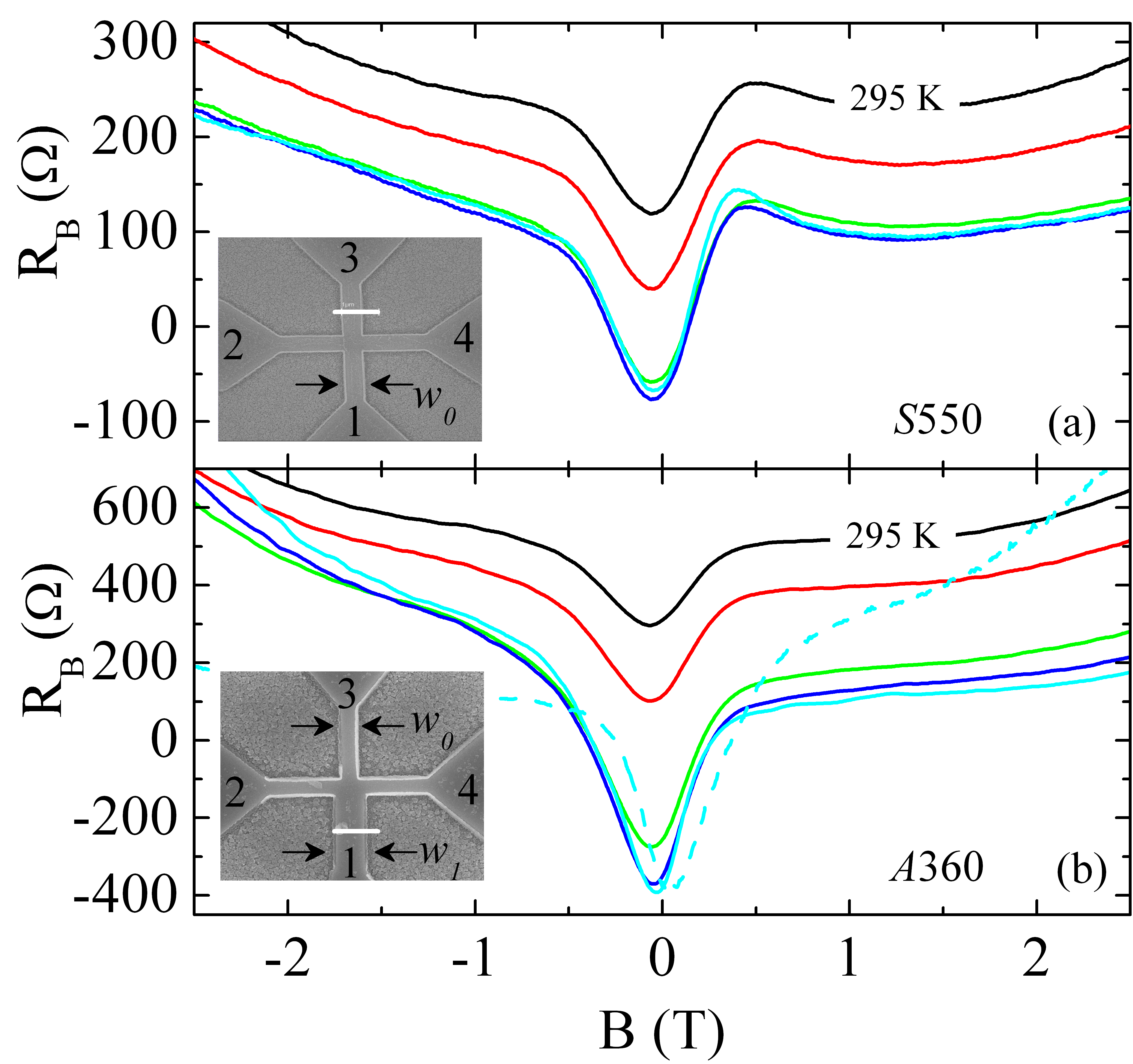}[bhp]
\caption{\label{}(a) and (b) Bend resistance R$_{B}$(B) = (V$_{4}$-V$_{3}$)/I$_{21}$ measurements at various temperatures between 160 K and 295 K for two different device geometries. From top to bottom T = 295 K, 280 K, 240 K, 200 K and 160 K. The dashed line in (b) represents the reciprocal measurement R$^{\dagger}_{B}$(B) at 160 K. Insets show electron micrographs of the devices. An AC current of 100 nA was used.}%
\end{figure}

We have investigated ballistic electron transport at elevated temperatures by studying the bend resistance R$_{B}$ = (V$_{4}$-V$_{3}$)/I$_{21}$ (see Fig. 1 insets). If at B = 0 a large proportion of electrons injected from lead 1 are ballistic, those with large forward momentum are transmitted directly to the opposite lead 3. This raises the potential of lead 3 with respect to lead 4, generating a negative bend resistance (NBR). A small magnetic field deflects the electron beam away from lead 3 into one of the side leads causing the NBR to decay. The resulting dip in the low-field R$_{B}$(B) (centered on B = 0) is a clear signature of ballistic transport. The NBR anomaly has previously not been observed above 200 K in InSb/AlInSb devices.

The bend resistance results obtained from a symmetric cross with physical lead width $w_{0}$ = 550 nm ($S$550) and an asymmetric cross with $w_{0} (w_{1}) = 360 (660)$ nm ($A$360) between 160 K and 295 K are shown in figure 1(a) and (b), respectively (see Fig.1 inset for device geometries). At 160 K, a distinct NBR is observed in both devices with an amplitude and full width at half maximum (FWHM) that remains approximately constant up to 240 K. Remarkably, the NBR feature persists up to 295 K indicating significant ballistic electron transport in the 2DEG. This result demonstrates that the parasitic effects of parallel conduction in the bottom growth buffer layer have been substantially suppressed by wafer design, and moreover that our processing strategy has not degraded the 2DEG mobility to the point that all carriers are diffusive as reported in InAs/AlSb heterostructures\cite{Folks2009}. The NBR feature is superposed on a background resistance (R$_{bg}$) that is approximately constant ($\approx 100 \Omega$) below 240 K and rises with increasing temperature such that R$_{B}$(0) is no longer negative. Nevertheless, ballistic coupling of leads 1 and 3 is clearly evident by the persistent dip at B = 0. This background will be discussed further in comparison to theoretical modeling. A secondary feature of the bend resistance data is the asymmetry of the field response. The geometrical origin of this is confirmed by measurements of the resistance R$^{\dagger}_{B}$ = (V$_{1}$-V$_{2}$)/I$_{34}$ which satisfies very closely the reciprocity relation R$_{B}$(B) = R$^{\dagger}_{B}$ (-B), as shown for device $A$360 by the dashed line in Fig. 1(b). 

\begin{figure}
\includegraphics[width=7.50cm]{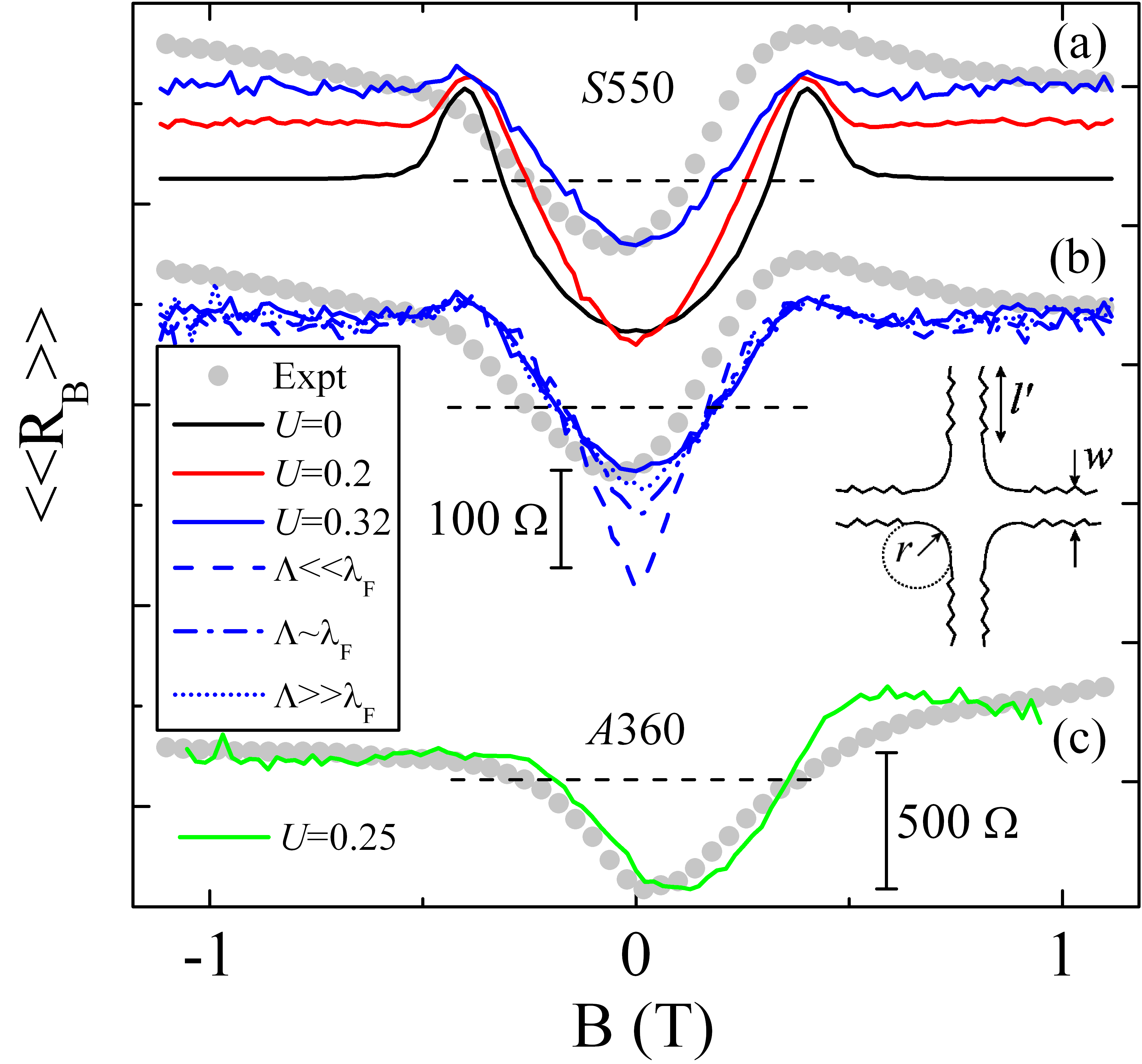}[htbp]
\caption{\label{} Quantum calculations of the energy-impurity averaged bend resistance $\ll$R$_{B}$(B)$\gg$ (lines) with varying disorder compared to experimental data at 160 K (symbols). (a) $S$550 with smooth sidewalls and impurity potential ($U$) as labeled. (b) $S$550 with $U$ = 0.32 and sidewall roughness as labeled. (c) $A$360 with $U$ = 0.25 and smooth sidewalls. Fluctuations are due to interference effects that are not fully averaged out.  Black dashed lines indicate R$_{B}$ = 0. [A corner rounding radius $r$ = 120 nm and lead length $l'$ = 1.7 $\mu$m was used in all cases, see inset].}%
\end{figure}

Measurements of the Hall resistance, R$_{H}$ = (V$_{4}$-V$_{2}$)/I$_{31}$, were also performed. The electron densities of the mesoscopic devices ($n_{mes}$) were estimated from the Hall slope, at field (B$\approx$ 0.5 T) where ballistic anomalies are absent, to be $n_{mes}$ = 5 (3.75) x 10$^{11}$ cm$^{-2}$ and $n_{mes}$ = 4.5 (3.5) x 10$^{11}$ cm$^{-2}$ at 295 K (160 K), for $S$550 and $A$360 respectively.

To gain further insight into the microscopic properties of the devices, we have performed extensive numerical quantum transport calculations for the bend resistance using a tight-binding code which combines the Green's function techniques of Baranger et al.\cite{Baranger1991} and Sanvito et al.\cite{Sanvito2008}. Some effects of finite temperature were simulated using a simple energy-averaging technique that takes into account the Fermi distribution.\cite{Baranger1991} We consider two possible types of disorder in the devices: elastic scattering from impurities, and from the sidewalls (shown to be important in our previous work)\cite{Gilbertson2011}. Impurity scattering was modeled with Anderson site-disorder: the on-site energies of the tight-binding Hamiltonian were chosen from an interval [-$U$, $U$] with uniform probability.\cite{Baranger1991} Sidewall scattering was taken into account by introducing a boundary roughness characterized by a mean amplitude ($\Delta$) and correlation length ($\Lambda$) after Akera et al.\cite{Akera1991}. For these calculations, we use $\Delta$ = 5 nm (deduced from atomic force microscopy of the lateral etched surface) and consider the three limits: $\Lambda << \lambda_{F}$, $\Lambda \approx \lambda_{F}$ and $\Lambda >> \lambda_{F}$, where $\lambda_{F} = 2\pi/k_{F}$ is the Fermi wavelength. Electron-phonon scattering is not included in this simple model, however, as we will show, the impurity scattering model successfully captures the essential features of momentum scattering within the channel and can be used to gain a realistic representation of the channel mobility.

Figure 2(a) and (b) shows the energy-impurity averaged bend resistance curves (lines), $\ll$R$_{B}$(B)$\gg$, for $S$550 with varying disorder compared to the experimental curve at 160 K symbols). Note that we have used experimenal $n_{mes}$ and the effective electrical lead widths $w_{eff}$ ($w_{eff}=w-2w_{dep}$). By comparison we can draw certain conclusions. In the absence of disorder (solid black line), $\ll$R$_{B}$(B)$\gg$ has zero background resistance, and an NBR around B = 0 that is both broader and larger in amplitude ($\ll\Delta$R$_{B}\gg$) than observed experimentally. Finite R$_{bg}$ is clear evidence of disorder in the experimental devices. The quantum calculations provide confirmation that the observed low-field ($<$ 1T) characteristics of R$_{B}$(B) are determined almost entirely by scattering in the channel rather than by scattering at the boundaries: ($i$) Calculations with only sidewall scattering yield $\ll\Delta$R$_{B}\gg$ up to 10 times greater than experimental data due to enhanced electron collimation\cite{Blaikie1992} (not shown for clarity); ($ii$) R$_{bg}$ is sensitive to the strength of impurity scattering [Fig. 2(a)] but is relatively insensitive to the presence, or type, of boundary roughness [Fig. 2(b)]; ($iii$) A comparison of $\ll\Delta$R$_{B}\gg$ with and without surface roughness [Fig. 2(b)] to experiment suggests little enhancement from diffuse collimation in the present devices. Indeed, good agreement with experimental data is found for the $\ll$R$_{B}$(B)$\gg$ curves with smooth boundaries ($\Delta$ = 0) and $U$ = 0.32 (solid blue lines). Similar results were found for device $A$360 [Fig. 2(c)]. Note that the assumption of smooth sidewalls is consistent with large $w_{dep}$ due to electrostatic screening of the exterior sidewall roughness.

The effective channel mobility ($\mu_{eff}$) for a given disorder can be obtained from calculations of G$_{2T}$ for single leads of varying length ($l'$) in the diffusive limit where the relation G$_{2T}$($l') = n_{mes}e\mu_{eff}(w/l')$ is valid. The disorder corresponding to the solid blue and green curves in Fig. 2(b) and (c) respectively, yield $\mu_{eff} \approx$ 45,000 cm$^{2}$/Vs. Comparing this value obtained from 0 K quantum calculations to that of the reference sample at 160 K, we find $\approx$ 30 $\%$ reduction in the former case, suggesting that some degradation has occurred due to the nanofabrication process.

\begin{figure}
\includegraphics[width=8.0cm]{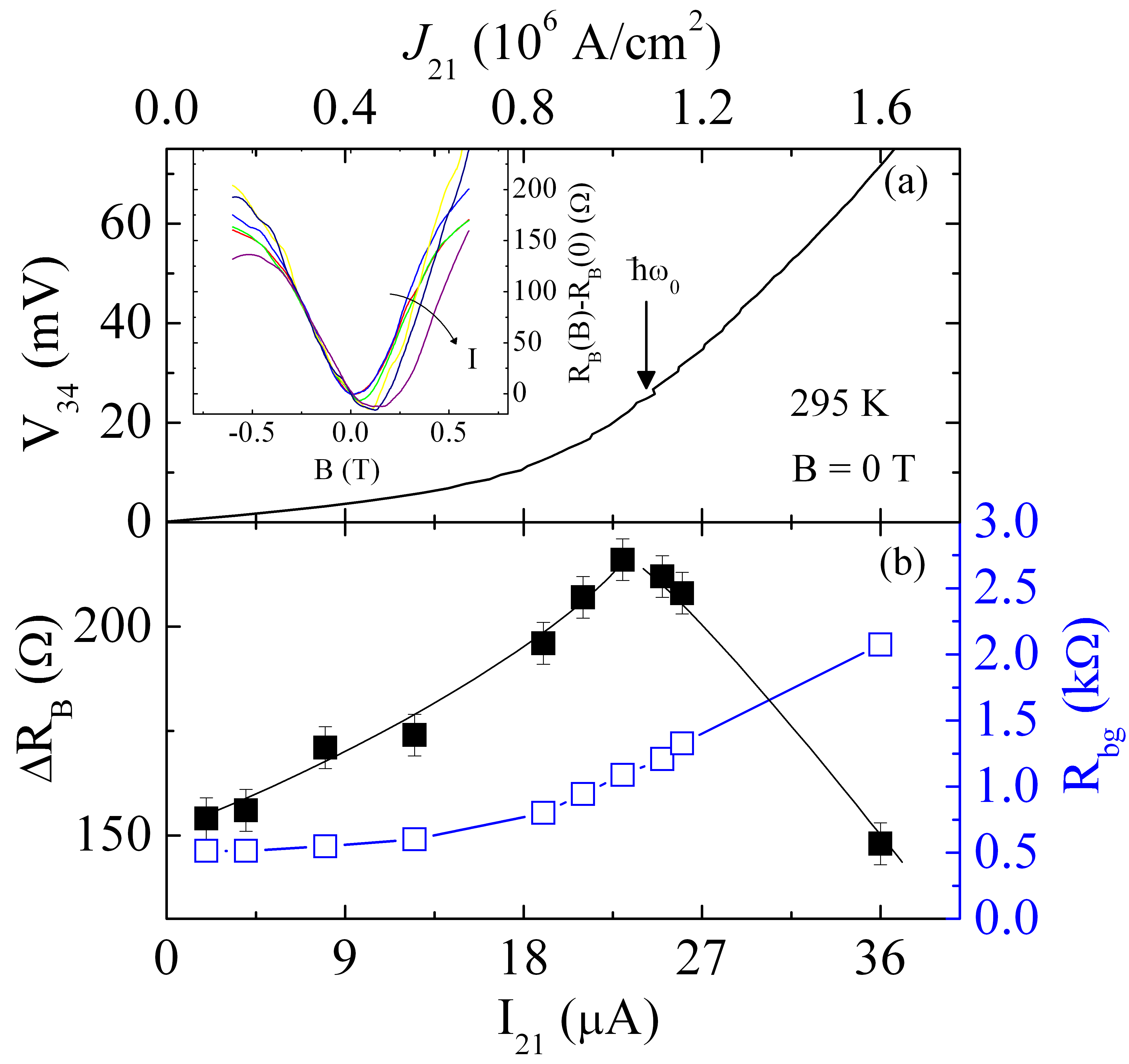}[htbp]
\caption{\label{}(a) IV characteristic of device $A$360 ($w_{0eff}$ = 130 nm) obtained at B = 0 T and 295 K. Inset: The DC bend resistance R$_{B}$(B) at various bias currents illustrating the effect of broadening. The zero field value is subtracted for clarity. (b) The dependence of the DC NBR amplitude (solid symbols) and background resistance (open symbols) on bias current. $\Delta$R$_{B}$(I$_{21}$) exhibits a turnover at I$_{21} \approx 25 \mu$A ($eV_{43} \approx \hbar \omega_{0}$) due to LO phonon emission that lowers $l_{0}$.}%
\end{figure}

We now discuss the operation of our devices in the high bias, nonequilibrium transport regime relevant for nanoelectronic applications where large signals are required. In principle hot-ballistic electron transport is limited by LO phonon emission ($\hbar\Omega_{0}=25$ meV in InSb), but at 295 K the thermal broadening of the electron distribution is already rather large ($\approx$ 26 meV) and thus hot electron effects should be less acute. 

The forward bias IV characteristic of device $A$360 ($w_{0eff} \approx$ 130 nm) at 295 K is shown in Fig. 3(a). A distinct nonlinearity is observed at high bias current, but note that the differential resistance is positive over the entire range due to the diffuse background. Figure 3(b) shows the non-monotonic dependence of the DC NBR amplitude $\Delta$R$_{B}$ on I$_{21}$. Synonymous with the above is a broadening of the NBR with increasing I$_{21}$ [see inset to Fig. 3(a)]. Since the FWHM (B$_{FWHM}$) is proportional to $k$, the broadening is a direct indication of the excess kinetic energy ($\Delta E \propto$ eV$_{34}$) gained in the nonequilibrium regime where $k \propto k_{F}(1+\Delta E/E_{F})^{1/2}$ and E$_{F}$ is the Fermi energy.\cite{Williamson1990} Likewise, if we assume $\Delta$R$_{B}$($\Delta E$,T) $\propto exp[-w_{0eff}/v_{F}(1+\Delta E/E_{F})^{1/2}\tau(\Delta E,T)]$,\cite{Schapers1995} where $v_{F}$ and $\tau(\Delta E$,T) are the Fermi velocity and momentum-scattering time respectively, the behavior for I$_{21} < 25 \mu$A can be understood by an increasing electron velocity, and $\tau(\Delta E$,T) that is essentially energy independent for $\Delta E < \hbar\Omega_{0}$. The turnover occurs at a voltage $e$V$_{34} \approx \hbar\Omega_{0}$ [see Fig. 3(a)], at which point injected electrons have sufficient energy to scatter by phonon emission, and $\tau(\Delta E,$T) is reduced substantially. This interpretation is consistent with the theory of optical phonon scattering, but contrary to previous reports at low temperature,\cite{Williamson1990,Schapers1995} the observed hot electron effects are considerably less acute and demonstrate that RT ballistic effects persist without decay up to $e$V$_{34}$ $\approx \hbar\Omega_{0}$. 

Finally, we can consider these devices as examples of quasi-ballistic Hall probes. The magnetic sensitivity is given by the noise-equivalent-field (NEF) B$_{NEF} = V_{n}$/I$_{31}R'$ where $V_{n}$ is the voltage noise and $R'$ the Hall coefficient ($\Omega$/T). For the device $A$360, we have R$_{2T}$ = 20 k$\Omega$ and $R'$ = 1390 $\Omega$/T at 295 K, giving B$_{NEF} \approx$ 500 nT/Hz for a bias current of 25 $\mu$A (where we have used the Johnson noise limit $V_{n}^{2}=4k_{B}TR_{2T}$). This sensitivity is considerably greater than previous reports of RT sub-micron Hall probes\cite{Folks2009,Kazakova2010} and magnetoresistance sensors,\cite{Solin2002,Boone2009} demonstrating that although not yet optimized, the ballistic cross structures we report here are highly competitive and hold significant promise for future high resolution RT sensors.

In summary, we have shown significant ballistic electron transport at RT in mesoscopic InSb QW devices. This result was achieved by using an epitaxial growth strategy that incorporated a PBL scheme to significantly reduce the effect of the thick buffer layer and the associated parasitic leakage current. Our results demonstrate that InSb QWs are a viable and promising alternative material for RT nanoelectronic applications.

This work was supported by the UK EPSRC under Grant No. EP/F065922/1. SAS is also supported by the US NSF under Grant No. ECCS-0725538, and NIH under Grant No. 1U54CA11934201, and has a financial interest in PixelEXX, a start-up company whose mission is to market imaging arrays.














%




%









%












\begin{thebibliography}{}

\bibitem{Javey2003} A. Javey, J. Guo, Q. Wang, M. Lundstrom, and H. Dai, Nature {\bf424}, 654 (2003).
\bibitem{Frank1998} S. Frank, P. Poncharal, Z. L. Wang, and W. A. d. Heer, Science {\bf280}, 1744 (1998).
\bibitem{Mayorov2011} A. S. Mayorov, $et al.$, Nano Letts. {\bf11}, 2396 (2011).
\bibitem{Song2001} A. M. Song, P. Omling, L. Samuelson, W. Seifert, I. Shorubalko, and H. Zirath, Appl. Phys. Letts. {\bf79}, 1357 (2001).
\bibitem{Song2001b} A. M. Song, P. Omling, L. Samuelson, W. Seifert, I. Shorubalko, and H. Zirath, Japanese J. Appl. Phys. {\bf40}, 909 (2001).
\bibitem{Hieke2000} K. Hieke and M. Ulfward, Phys. Rev. B {\bf62}, 16727 (2000).
\bibitem{Orr2008} J. M. S. Orr, A. M. Gilbertson, M. Fearn, O. W. Croad, C. J. Storey, L. Buckle, M. T. Emeny, P. D. Buckle, and T. Ashley, Phys. Rev. B {\bf77}, 165334 (2008).
\bibitem{Goel2004} N. Goel et al., Physica E {\bf20}, 251 (2004).
\bibitem{Gilbertson2011b} A. M. Gilbertson, P. D. Buckle, M. T. Emeny, T. Ashley, and L. F. Cohen, Phys. Rev. B {\bf84}, 075474 (2011).
\bibitem{Folks2009} L. Folks and et al., J. Phys.: Cond. Matt. {\bf21}, 255802 (2009).
\bibitem{Baranger1991} H. U. Baranger, D. P. DiVincenzo, R. A. Jalabert, and A. D. Stone, Phys. Rev. B {\bf44}, 10637 (1991).
\bibitem{Sanvito2008} S. Sanvito, C. J. Lambert, J. H. Jefferson, and A. M. Bratkovsky, Phys. Rev. B {\bf59}, 11936 (1999); I. Rungger and S. Sanvito, Phys. Rev. B {\bf78}, 035407 (2008).
\bibitem{Gilbertson2011} A. M. Gilbertson, M. Fearn, A. Kormanyos, D. E. Read, C. J. Lambert, M. T. Emeny, T. Ashley, S. A. Solin, and L. F. Cohen, Phys. Rev. B {\bf83}, 075304 (2011).
\bibitem{Akera1991} H. Akera and T. Ando, Phys. Rev. B {\bf43}, 11676 (1991).
\bibitem{Blaikie1992} R. J. Blaikie, K. Nakazato, J. R. A. Cleaver, and H. Ahmed, Phys. Rev. B {\bf46}, 9796 (1992).
\bibitem{Williamson1990} J. G. Williamson, H. van Houten, C. W. J. Beenakker, M. E. I. Broekaart, L. I. A. Spendeler, B. J. van Wees, and C. T. Foxon, Phys. Rev. B {\bf41}, 1207 (1990).
\bibitem{Schapers1995} T. Schapers, G. J. Kruger, J. Appenzeller, A. Forster, B. Lengeler, and H. Luth, Appl. Phys. Lett. {\bf66}, 3603 (1995).
\bibitem{Kazakova2010} O. Kazakova, V. Panchal, J. Gallop, P. See, D. C. Cox, M. Spasova, and L. F. Cohen, J. Appl. Phys. {\bf107}, 09E708 (2010).
\bibitem{Solin2002} S. A. Solin, D. R. Hines, A. C. H. Rowe, J. S. Tsai, Y. A. Pashkin, S. J. Chung, N. Goel, and M. B. Santos, Appl. Phys. Letts. {\bf80}, 4012 (2002).
\bibitem{Boone2009} T. D. Boone, N. Smith, L. Folks, J. A. Katine, E. E. Marinero, and B. A. Gurney, IEEE {\bf30}, 117 (2009).

\end{thebibliography}

\end{document}